# SUPERGRAVITY WAS DISCOVERED BY D.V. VOLKOV AND V.A. SOROKA IN 1973, WASN'T IT?

Steven Duplij

*Center for Information Technology (ZIV), Westfälische Wilhelms-Universität Münster, Münster, Germany*
*E-mail: douplii@uni-muenster.de*

Supergravity is a remarkable theory combining supersymmetry and general relativity. While the theory has many developers from across the globe, we wish to address the question who were the real originators of this fantastic idea.

The idea of supergravity (without mentioning this beautiful word) was given to the western public on December 5, 1972, in the last paragraph of the paper D.V. Volkov (47 y.o. that time, Dr. Habilitation, second degree after Ph.D. in European countries) and his PhD student V.P. Akulov (28 y.o. that time) "Possible universal neutrino interaction" ZhETF Pis. Red. (JETP Letters translated into English by AIP [1]) followed by the paper Physics Letters, September 3, 1973 "Is the neutrino a Goldstone particle?" [2]. It was clearly (for professionals) written: "...the gravitational interaction may be included by means of introduction the gauge fields for the Poincaré group. ...if the gauge field for the transformation (3) is also introduced, then as a result of the Higgs effect the massive gauge field with spin three-half appears and the considered Goldstone particle with spin one half disappears."

This program was realized as a concrete model in 1973 by D.V. Volkov and his PhD student V.A. Soroka (29 y.o. that time) in the paper "Higgs effect for Goldstone particles with spin 1/2" ZhETF Pis. Red. (JETP Letters translated into English by AIP [3]).

Therefore, October 20, 1973, is the day of true discovery of supergravity as a physical model (in a nonlinear realization). The later publications can be considered only as further development of the same idea, in either a different realization or in modern terminology, which is, in general, a matter of taste.

Indeed, such a model (among numerous ones suggested later on) appeared in 1976 and contained a new "magic" word "supergravity" which was absent in the above Volkov-Soroka paper: "Progress toward a theory of supergravity" Physical Review D [4] by D. Z. Freedman (33 y.o. in 1972, a postdoc working on Regge poles and scattering, that time), P. van Nieuwenhuizen (34 y.o. in 1972, a postdoc working on muon scattering, that time), and S. Ferrara (27 y.o. in 1972, PhD working on conformal invariance, that time). Because the F-N-F paper cites the Volkov-Soroka article [3], the F-N-F paper can be considered as "discovering" only a new word "supergravity". In the same month (June, 1976), the paper "Consistent supergravity" Physics Letters B by S. Deser (45 y.o., in 1976, a professor working on quantum gravity and strings) and B. Zumino (53 y.o., that time, a professor working on supergauge theories) was published [5]. This paper cited the F-N-F article as a preprint and also V. Akulov, D.V. Volkov, and V.A. Soroka paper of 1975 "Gauge Fields on Superspaces with Different Holonomy Groups" [6] which contains the reference to Volkov-Soroka (1973) [3].

The history of supergravity is given in the "SUSY story (narrated by its founders)" [7] and in the article "Supergravity" [8] in the "*Concise Encyclopedia of Supersymmetry*" by S. Duplij, J. Bagger, W. Siegel (Eds.) [9].

The connection between the V-S and F-N-F approaches was clearly explained several times, e.g.:

1) by D.V. Volkov in the above Encyclopedia "Supergravity before and after 1976. The story of goldstonions" or here [10] and in his talk "Supergravity before 1976" at the *International Conference on the History of Original Ideas and Basic Discoveries in Particle Physics,* Erice, 1994 [11];

2) by V.A. Soroka in "The Sources of Supergravity" in "*The Supersymmetric World. The Beginnings of the Theory*" G. Kane and M. Shifman (Eds.) [12] or [13] and "Starting-point of Supergravity" [14] presented at the Conference "Supergravity at 25", Stony Brook NY, 2001.

It is remarkable and a pity, that in the mass media articles about numerous prizes given for the "discovery of supergravity" contain not a single word about the true discoverers of supergravity in 1973: Dmitriy Vasilievich Volkov (1925-1996, Memorial page [15]) and Vyacheslav Alexandrovich Soroka (1944-2011, Memorial page [16]).

**ORCID IDs**

Steven Duplij https://orcid.org/0000-0003-1184-6952

**СУПЕРГРАВІТАЦІЯ БУЛА ВІДКРИТА Д.В. ВОЛКОВИМ ТА В.А. СОРОКОЮ У 1973 РОЦІ, ЧИ НЕ ТАК?**
Степан Дуплій

*Центр інформаційних технологій, Вестфальський університет Вільгельма, Мюнстер, Німеччина*

Супергравітація – чудова теорія, яка об'єднує суперсиметрію і загальну теорію відносності. Хоча цією теорією займалося багато вчених у всьому світі, ми намагаємося з'ясувати, хто в реальності відкрив цю фантастичну ідею.
**КЛЮЧОВІ СЛОВА**: супергравітація, суперсиметрія, голдстоніан, ефект Хіггса, група Пуанкаре, калібрувальне поле

**СУПЕРГРАВИТАЦИЯ БЫЛА ОТКРЫТА Д.В. ВОЛКОВЫМ И В.А. СОРОКОЙ В 1973 ГОДУ, НЕ ТАК ЛИ?**
Степан Дуплий

*Центр информационных технологий, Вестфальский университет Вильгельма, Мюнстер, Германия*

Супергравитация – замечательная теория, объединяющая суперсимметрию и общую теорию относительности. Хотя этой теорией занимались многие ученые во всем мире, мы пытаемся выяснить, кто в реальности открыл эту фантастическую идею.
**КЛЮЧЕВЫЕ СЛОВА:** супергравитация, суперсимметрия, голдстониан, эффект Хиггса, группа Пуанкаре, калибровочное поле